\documentclass[conference]{IEEEtran}
\IEEEoverridecommandlockouts
\usepackage{cite}
\usepackage{amsmath,amssymb,amsfonts}
\usepackage{algorithmic}
\usepackage{graphicx}
\usepackage{textcomp}
\usepackage{xcolor}
\usepackage{url}
\usepackage{hyperref}
\usepackage[a4paper, total={184mm,239mm}]{geometry}
\def\BibTeX{{\rm B\kern-.05em{\sc i\kern-.025em b}\kern-.08em
    T\kern-.1667em\lower.7ex\hbox{E}\kern-.125emX}}
\begin{document}

\title{OpenSerDes: An Open Source Process-Portable All-Digital Serial Link
}
\author{\IEEEauthorblockN{}
\IEEEauthorblockA{\textit{} \\
\textit{}\\
 \\
}}

\author{\IEEEauthorblockN{Gaurav Kumar K, Baibhab Chatterjee and Shreyas Sen}
\IEEEauthorblockA{School of ECE, Purdue University, West Lafayette, USA-47906. email:\{gauravk, bchatte, shreyas\}@purdue.edu}}

\maketitle

\begin{abstract}
Over the last decade, the growing influence of open source software has necessitated the need to reduce the abstraction levels in hardware design. Open source hardware significantly reduces the development time, increasing the probability of first-pass success and enable developers to optimize software solutions based on hardware features, thereby reducing the design costs. The recent introduction of open source Process Development Kit (OpenPDK) by Skywater technologies in June 2020 has eliminated the barriers to Application-Specific Integrated Circuit (ASIC) design, which is otherwise considered expensive and not easily accessible. The OpenPDK is the first concrete step towards achieving the goal of open source circuit blocks that can be imported to reuse and modify in ASIC design. With process technologies scaling down for better performance, the need for entirely digital designs, which can be synthesized in any standard Automatic Place-and-Route (APR) tool, has increased considerably, for mapping physical design to the new process technology. This work presents a first open source all-digital Serializer/Deserializer (SerDes) for multi-GHz serial links designed using Skywater OpenPDK 130nm process node. To ensure that the design is fully synthesizable, the SerDes uses CMOS inverter based drivers at the transmitter, while the receiver front end comprises a resistive feedback inverter as a sensing element, followed by sampling elements. A fully digital oversampling CDR at the receiver end recovers the transmitter clock for proper decoding of data bits. The physical design flow utilizes OpenLANE, which is an open source end-to-end tool for generating GDS from RTL. Cadence Virtuoso has been used for extracting parasitics for post-layout simulations, which exhibit the SerDes functionality at 2 Gbps for 34 dB channel loss while consuming 438 mW power. The generated GDS and netlist files of the SerDes, along with the required documentation, are uploaded in a GitHub repository for public access.\color{black}
\end{abstract}


\section{Introduction}
\subsection{The Need for open source Hardware}
With the predominance of open source software in recent times, the term “open source hardware”, introduced by Bruce Perens in 1997, is gaining momentum in the design ecosystem \cite{Tim_Ansel, 
Why_open_hardware}. Open source hardware refers to a hardware, an ASIC, or even, a circuit  that  has  provided access to its entire design, specifications, and documentations, which  can  be  used,  altered,  or  distributed  by  anyone.  Like source code in case of open source software, all the schematics, logic  designs,  layout-data, and netlists  need  to  be  made  available for  revisions  by  anyone,  who  has  access  to
\begin{figure}[htp]
   \centering
    
    \includegraphics[width=0.48\textwidth]{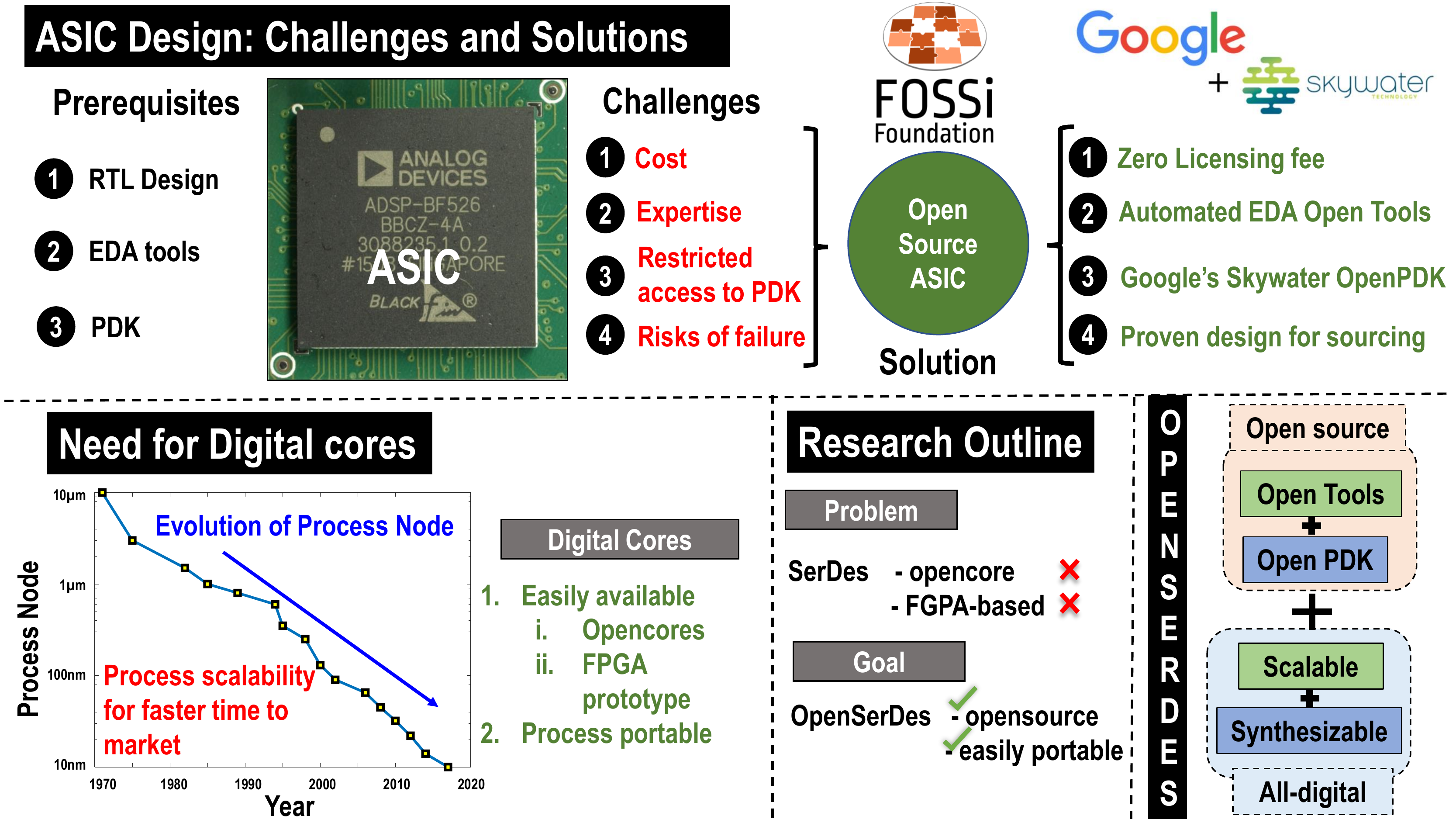}
    \caption{Motivation for open source hardware design and OpenSerDes}
    \label{fig:motivation}
    \vspace*{-0.2in}
\end{figure}
the  tools  to  read, manipulate  and  update  the  existing  design  with  new  features, usually  aiming  for 
better  performance  and  share  the improved design back to the community for further enhancements/evaluations. 

Open source hardware reduces the design time, enables software developers to effectively optimize their source codes based on hardware features, accelerates innovations with multiple people working together and sharing their knowledge, reducing the cost. This illustrates the need for open source hardware for faster technological advancement.

Several Open source hardware projects, like RISC-V\cite{RISC_V_repo}, UC Berkley’s BAG  \cite{BAG_repo}, Google’s OpenTitan\cite{OpenTitan_repo}, OpenRoad\cite{Open_road_repo} etc., have started uploading their design specifics in the repositories for public use. Berkeley Analog Generator (BAG) \cite{BAG_CICC} is an open source tool that provides framework for generating Analog and Mixed Signal (AMS) circuits from the design specifications. Numerous Open source tools like magic, ngspice, etc. have been used extensively for design layout and simulations. The Free and Open Source Silicon (FOSSi) foundation promotes free and open digital hardware designs, supporting open standards development. The recent introduction of Skywater 130 nm open source PDK\cite{Tim_Ansel, Open_PDK_repo} by Google and Skywater Foundries, has opened several opportunities for circuit designers. The conventional PDKs from TSMC, Global Foundries, UMC etc. are quite expensive and involve strict licensing process, making them difficult to access for all. OpenPDK has removed the roadblocks for ASIC designing, providing access to make custom chips for free initially for faster development cycle or at very low-cost eventually. 

With the process technologies scaling down for better performance, the existing designs or circuits need to be ported to latest process, reinstating the fact that “process-portability” of the design plays a key role in selecting the circuit topology. Although Analog/Mixed signal circuit design approach may exhibit superior performance than constituent synthesizable digital-only circuits, in terms of speed, power, etc., the entire design needs to be redrawn from scratch, to port to a new process. This promotes for synthesizable, digital-friendly designs that can be easily ported from one process to another. 

SerDes, an important building block in communication systems, helps in transmitting a parallel stream of data from block to another over serial links, thereby reducing the number of I/O pins, device dimensions, and power consumption of the system. The unavailability of opencore or FPGA-based digital IO/SerDes prevents faster time-to-market for designs, as they demand considerable time and effort for porting across technologies. This work presents the functionality of fully open source, multi-GHz, all-digital SerDes link, built using Skywater 130nm Open PDK, ensuring process-portability, thereby reducing the overhead porting time.



\subsection{Our Contribution}
\begin{itemize}
    \item This work presents the first open source \textbf{all-digital SerDes}, with multi-GHz capabilities. Unlike traditional mixed-signal SerDes designs, the all-digital design makes the present design \textbf{process-portable}, which is a key aspect of \textbf{Open-Source Hardware.}
    \item This work uses, for the first time, an \textbf{open source} PDK, skywater 130 nm, for the realization of \textbf{SerDes}.
    \item The work illustrates the importance of open source hardware and synthesizable design techniques, motivating their adoption in circuit design.

\end{itemize}

\subsection{Paper Organization}
The remainder of the paper is organized as follows. Section II illustrates the importance for open source hardware by comparing with traditional design techniques. In Section III, the typical architecture of the SerDes link is demonstrated, discussing the various design considerations, while promoting the need for all-digital design. Section IV presents the architecture of the automated SerDes design. Section V discusses the implementation results of the SerDes. Section VI explains the pros and cons of the design with use-cases and design flow of the OpenLANE tool. We conclude the paper in Section VII.

\section{open source hardware vs. Traditional Design}
Open hardware removes the common hindrances to system design and ensures faster development time. Hardware/ASIC designers can reduce the risk of design failure by leveraging pre-existing designs in an open source hardware library, which enables a “plug and play” feature for previously proven successful hardware constructs, which was not possible earlier\cite{Why_open_hardware}.
\begin{figure}[htp]
   \centering
    \includegraphics[width=0.45\textwidth]{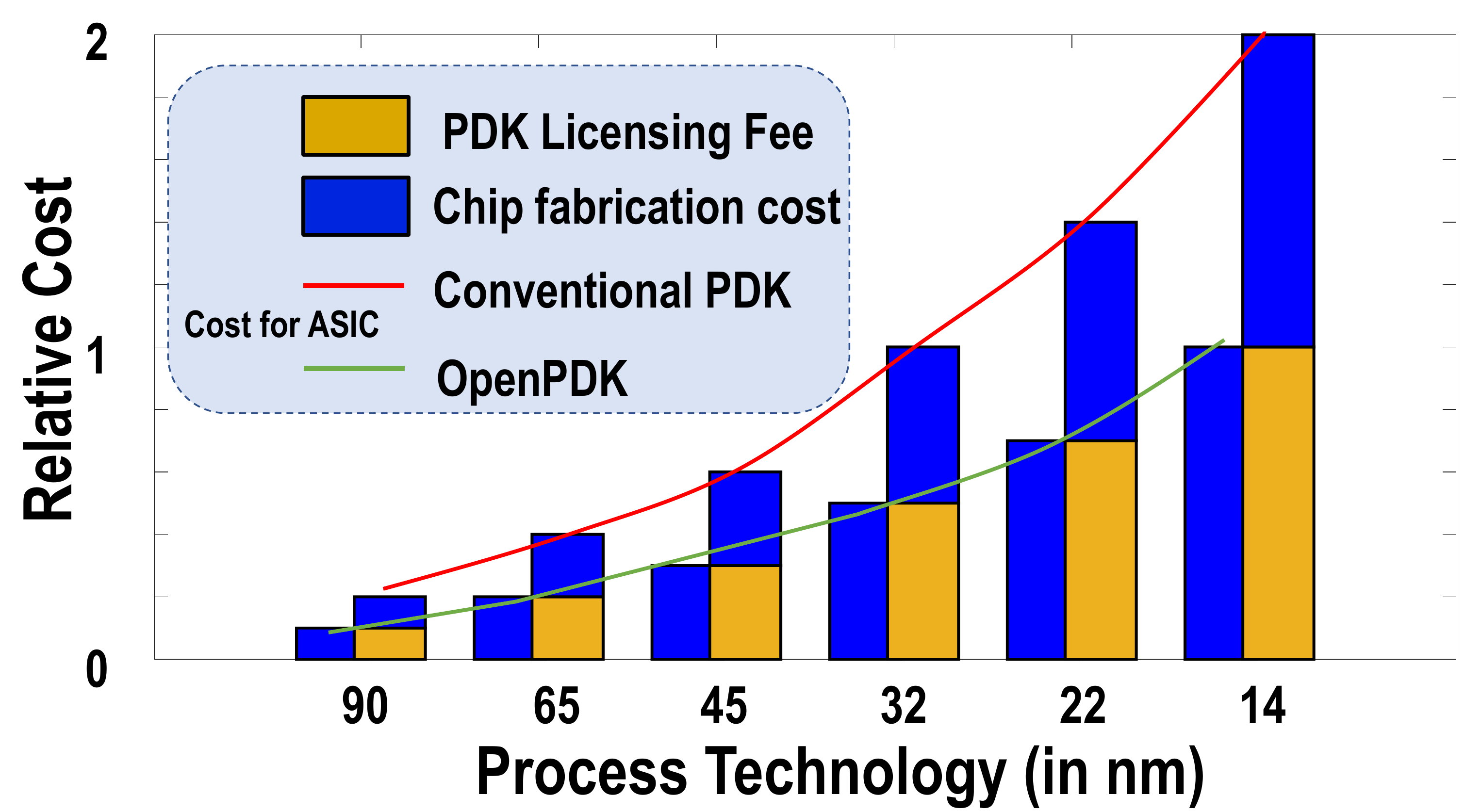}
    \caption{Comparison of Relative chip fabricating cost using Open source PDK and traditional PDKs \cite{Chip_Cost}}
    \label{fig:Open_vs_Non_open}
    \vspace*{-0.2in}
\end{figure}
Designers possess the complete design that can be modified based on their required specifications. This prevents re-inventing the designs, enables rapid innovation and progress, while lowering the design cost. Figure \ref{fig:Open_vs_Non_open} shows the comparison between the costs incurred in manufacturing chip using traditional PDKs (i.e., TSMC, GF, Intel, etc.), and skywater 130nm openPDK. It can be seen that openPDK reduces the PDK licensing cost, bringing down the overall chip manufacturing cost. Note that the PDK licensing cost is not publicly accessible and thus, it has been scaled relative to fabrication cost and process technology{\cite{Chip_Cost}}.

The involvement of community facilitates faster evolution and reception of the design\cite{Why_open_hardware_EDN}. It enables people to interact and share their knowledge of hardware design, thereby opening more avenues for innovation. It also provides the software developers the ability to optimize and enhance their solutions based on the hardware features, impacting the overall performance of the system. Ultra-low power pervasive IoT, being at the juncture of Moore's Law and Shannon's law, calls for innovation and collaboration at both software and hardware fronts \cite{DnT_Baibhab}, and hence would benefit immensely from open-source hardware designs.

Open source hardware has started to gain prominence in the design space, with numerous open source projects, like  BAG, RISC-V, OpenTitan, OpenPDKs etc., promoting open standard developments. In a nutshell, open source hardware provides people with a platform to share their knowledge and determine the course of the technology. 
\begin{figure*}[t]
  \centering
  \includegraphics[width=0.85\textwidth]{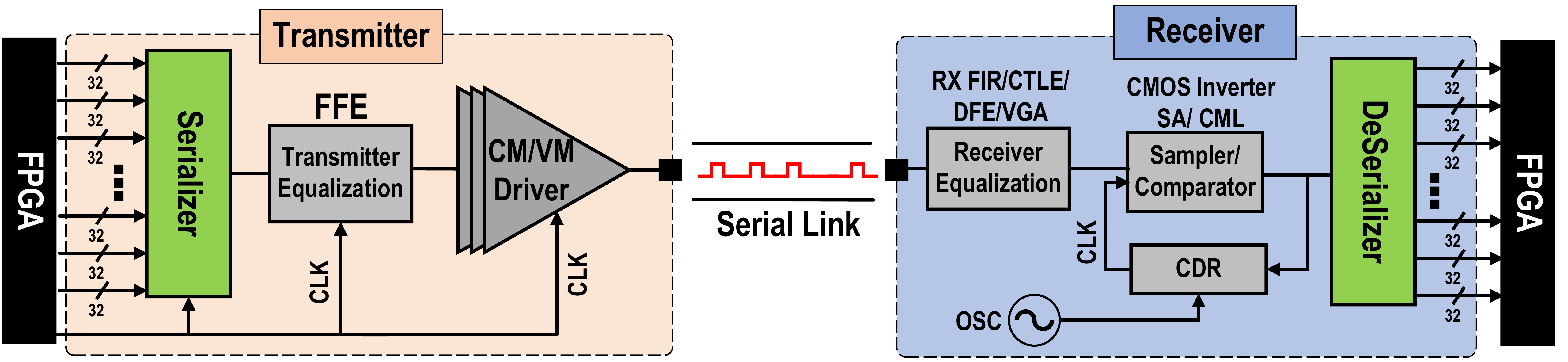}
  \caption{Block-level overview of a typical SerDes Architecture}
  \label{Serdes_overview}
  \vspace*{-0.2in}
\end{figure*}
\section{SerDes Design Considerations}

SerDes has emerged as a popular solution in communication system design since the paradigm shift away from traditional parallel communication standards to overcome the voltage scaling dependencies and performance limitations. Serial links, unlike parallel implementations, are scalable and require fewer I/Os, achieve smaller device footprints, thereby having lower power requirements.
The growing demand for low-power high-speed link has motivated development of innovative circuit techniques for SerDes over the last few years\cite{historical_trends_razavi, Prof_32, Prof_Chintan}. Mixed-signal designs, having multiple feedback loops from analog and digital domains, have presented a cost-effective solution, reaping the benefits of both analog and digital worlds. But with consequent technology scaling, requiring faster time-to-market for lower process node designs, there is a need for all-digital SerDes architecture, which can be easily ported across technologies.
 


Figure \ref{Serdes_overview} shows the block-level architecture of a typical SerDes. The transmitter (TX) front end comprises of a serializer, TX equalization block and a driver, while the Receiver (RX) front end has RX equalization, sampling element and a deserializer\cite{RF_Micro}. In case of non-forwarded-clock architecture, where timing signal of the transmitter, i.e. clock, is not transmitted along with data, a Clock and Data Recovery (CDR) circuit recovers the timing information and facilitates correct decoding of received data.

TX Feed Forward Equalizer (FFE), used as TX equalization block, pre-distorts the data over some bit durations to negate the effect of channel loss/distortion. 
Transmit drivers, implemented either in Voltage Mode (VM) or Current Mode (CM) using differential amplifiers, provide required output voltage swing and match the channel impedance to reduce reflections. 
RX equalization improves signal gain and input referred noise of the received data. It is achieved using a Continuous Time Linear Equalizer (CTLE), a Decision Feedback Equalizer (DFE) or FIR filter based FFE\cite{RF_Razavi}.
Comparator/Sampler blocks resolve the data to logic levels based on some threshold or differential value. 
They can be realized using static amplifiers or clocked regenerative amplifiers. Regenerative comparators, such as a Strong Arm Latch \cite{SA_razavi} samples the input at clock edges and resolves the differential to logic levels by positive feedback based latching action. Recently, high speed ADCs are used as comparators to generate digital data from received signal\cite{ADC_based_112Gb_ISSCC_2020}.
A CDR extracts the timing information from the incoming data transitions, by adjusting the phase of the sampling clock, and uses this recovered clock to resample the data and decode the data correctly. 

Our earlier works \cite{Prof_32} \cite{Prof_Chintan} demonstrate high speed bi-directional electrical link architecture, with TX and RX Equalization, for Proximity communication in 22nm and 14nm CMOS technologies. However, these designs consist analog/mixed-signal circuits for higher performance, and hence is not amenable to a fully digital implementation.

\subsection{Need for All-digital design}
Although mixed signal systems offer greater speeds, higher sensitivity, and better power efficiency than both analog and digital counterparts, the push for an all-digital design is garnering much interest. As the technology scales down to a new one, digital designs ensure faster time to market than analog or mixed signal designs because significant amount of time is spent on refabricating the entire analog design from scratch with the updated design rules of new technology, whereas in case of all-digital designs, the complete circuitry can be obtained by running the RTL through Synthesis and APR tool that maps the design to latest technology, abiding all the latest design rules. This saves considerable amount of man-hours. Innovations in technology scaling is primarily focused on the performance of digital systems, compromising the performance of analog circuitry, because of the occurrence of various non-ideal effects and increased variability in MOS device, in such smaller dimensions and lower supply voltage regime. Fully synthesizable designs help in reaping  full benefits from new technology, where lower voltage supplies limit the circuit and performance feasibilities of analog circuits.

Note that all-digital designs are well suited in scenarios where the possibility of technology scaling is imminent (in case of  Skywater 130 nm OpenPDK).  The main challenge is to maximize the performance of all-digital design, in terms of speed, sensitivity and power consumption, to match up to the standards of mixed signal design.



\section{Automated SerDes Design}
An all-digital SerDes is generated using the OpenLANE Tool\cite{Openlane_repo}, which provides an open source environment for automatic RTL to GDS generation without manual intervention. The primary goal is to ensure that the design is fully synthesizable, without compromising the performance of the SerDes, making it easier to port from one process technology to another. The design comprises of a serializer and CMOS inverter based transmit driver as a transmitter, a resistive-feedback inverter-based \cite{Res_FB_inverter} sensing element and a CMOS inverter to generate rail-to-rail signal, with a flip-flop-based sampling element, followed by the deserializer, as receiver. An oversampling\cite{OverSamp_Overview} \color{black} CDR facilitates the proper detection of data by generating correct timing information based on received data. The complete design is available at: \url{https://github.com/SparcLab/OpenSERDES}.


\subsection{Transmitter Architecture}
The transmitter converts the input parallel data streams into serial bits, before transmitting them. It comprises of serializer followed by driver to drive the output voltage to the channel.

\paragraph {Serializer}
The Serializer block is coded in Verilog HDL and synthesized using OpenLANE tool to generate the GDS, mapped to Skywater 130nm technology. Considering realistic scenarios, the serializer is designed to take in 8 parallel input data streams of 32 bits each and produces serial bits. It is implemented using a Finite State Machine (FSM) that sequentially takes one parallel data stream and serializes them.
\begin{figure}[htp]
   \centering
    \includegraphics[width=0.47\textwidth]{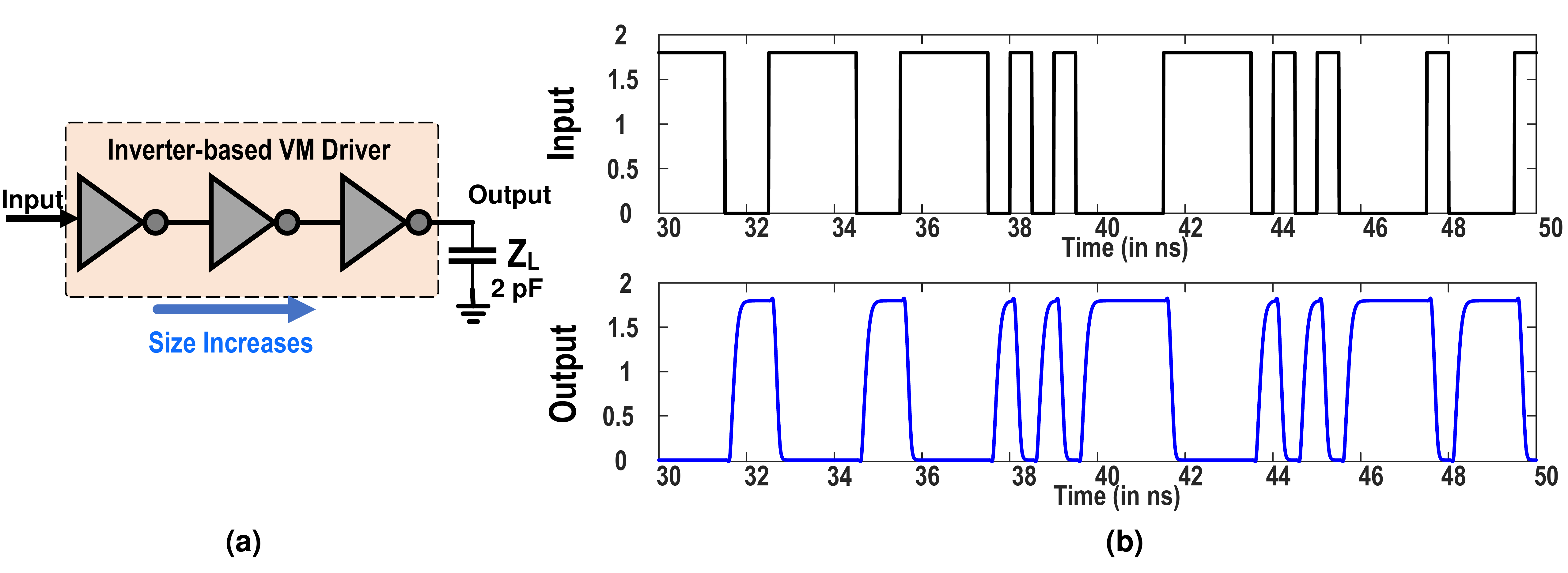}
    \caption{(a) Schematic of CMOS Transmit Driver (b) Input/Output Waveforms of Driver at 2 Gbps with 2 pF termination }
    \label{fig:Transmit_Driver}
\end{figure}

\paragraph{Transmit Driver}
A voltage-Mode (VM) CMOS driver is chosen provide rail-to-rail input swing at channel input. VM drivers have an advantage of reduced driver power, as compared to current-mode (CM) drivers. A three-stage inverter chain is used to implement the driver, to achieve the driving capability of 2 pF. Note that the individual CMOS stages are sized appropriately to obtain area and power optimal design. Figure \ref{fig:Transmit_Driver} shows the schematic representation of designed CMOS inverter-based driver, with related input-output waveforms. 

\subsection{Receiver Architecture}
\begin{figure}[htp]
   \centering
    \includegraphics[width=0.48\textwidth]{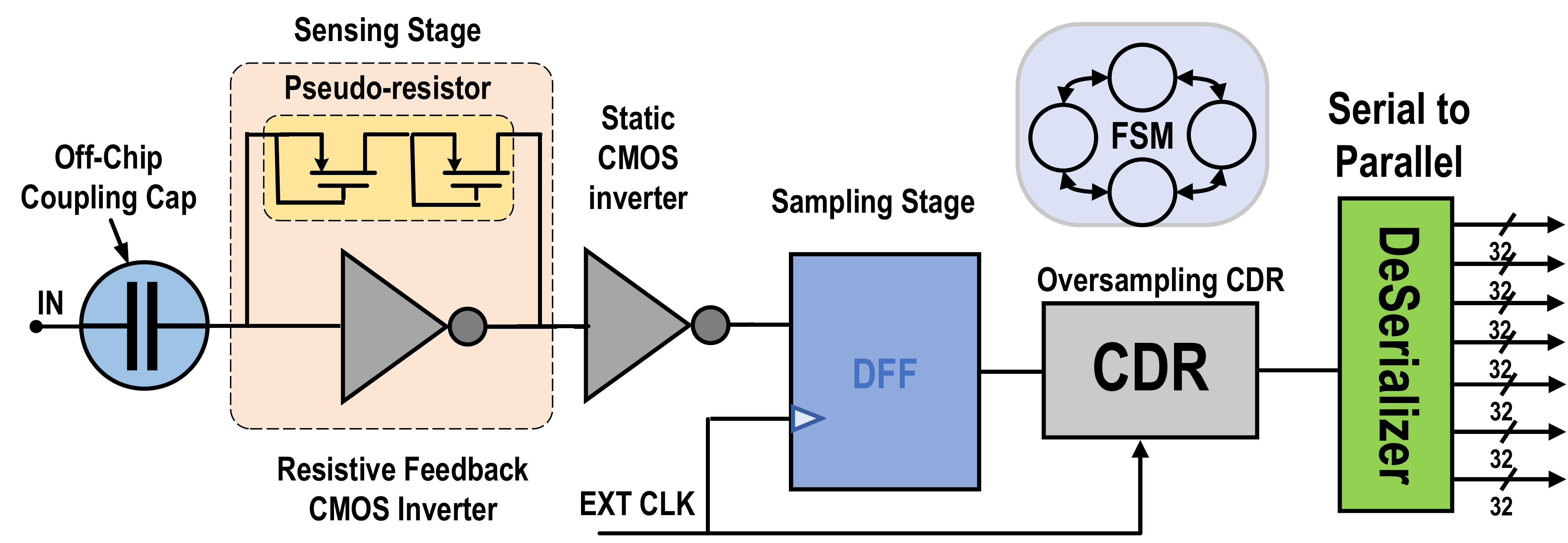}
    \caption{Receiver architecture of presented SerDes}
    \label{fig:Receiver_Architecture}
\end{figure}

The receiver front end uses a resistive feedback inverter as gain and sensing element and a CMOS inverter based static comparator for sampling the received signal. These blocks are chosen to ensure that the SerDes receiver is synthesizable and achieves desired performance criteria. Figure \ref{fig:Receiver_Architecture} presents the complete receiver architecture of the SerDes.

\begin{figure}[htp]
   \centering
    \includegraphics[width=0.48\textwidth]{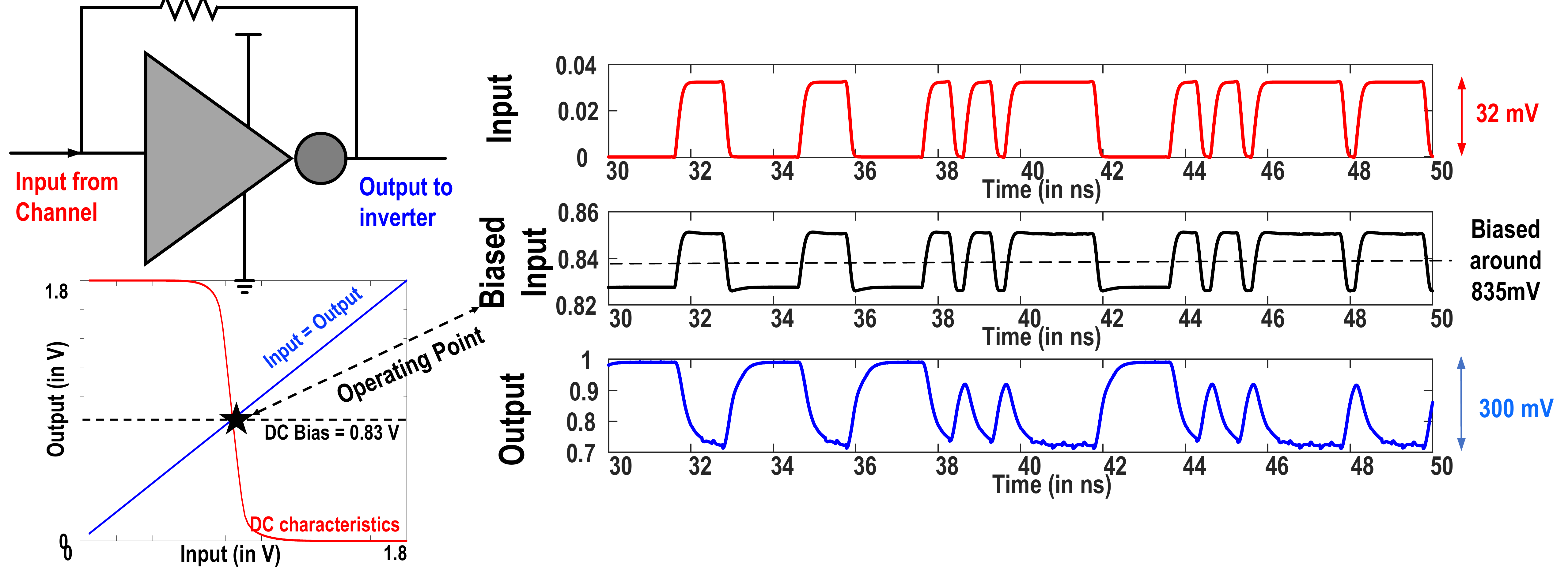}
    \caption{(a) Operating point, (b) Input/Output Waveforms, of Resistive feedback inverter}
    \label{fig:Res_FB_inv}
\end{figure}

\paragraph {Resistive Feedback Inverter}
The Resistive feedback inverter\cite{Res_FB_inverter} is a self-biased CMOS inverter with a resistive feedback from output to input, aiming to bias the inverter near the switching threshold, i.e. 0.5V$_{DD}$ point, enabling the CMOS inverter to work on signals with amplitudes much lower than its threshold voltage. To ensure synthesizability, the feedback resistor is realized as pseudo resistor using PMOS, where the gate and source are shorted to operate in linear resistive region. An off-chip capacitor is used to AC couple the received signal and prevent the received signal to interfere with the generated self-bias voltage, thereby maintaining the DC bias. Since the inverter operates at the point where both PMOS and NMOS are in saturation, there always exists a path for current to flow from supply to ground, resulting in non-zero static power disspiation. Figure \ref{fig:Res_FB_inv} shows the operating point of the designed self-bias inverter, with waveforms depicting its functionality.

\paragraph{Sampling Block}
The sampling block consists of CMOS inverter, which provides rail-to-rail signal at the input of D flip-flop, which then samples and forwards the received data to the deserializer, using the timing information provided by clock signal. Although the realized sampling topology has limited gain as compared to regenerative samplers/comparators, the front-end resistive feedback inverter compensates and provides the necessary gain.   

\paragraph{Deserializer}
The deserializer block converts serial bits into parallel data streams, which are to be interfaced with other blocks. It is implemented through an FSM that takes in serial bits and produces 8 parallel data streams of 32 bits each at the end of its operation. The entire design is synthesized using OpenLANE tool to generate the GDS and netlists. 

\subsection{Oversampling \color{black} CDR}
\begin{figure}[htp]
   \centering
\includegraphics[width=0.48\textwidth]{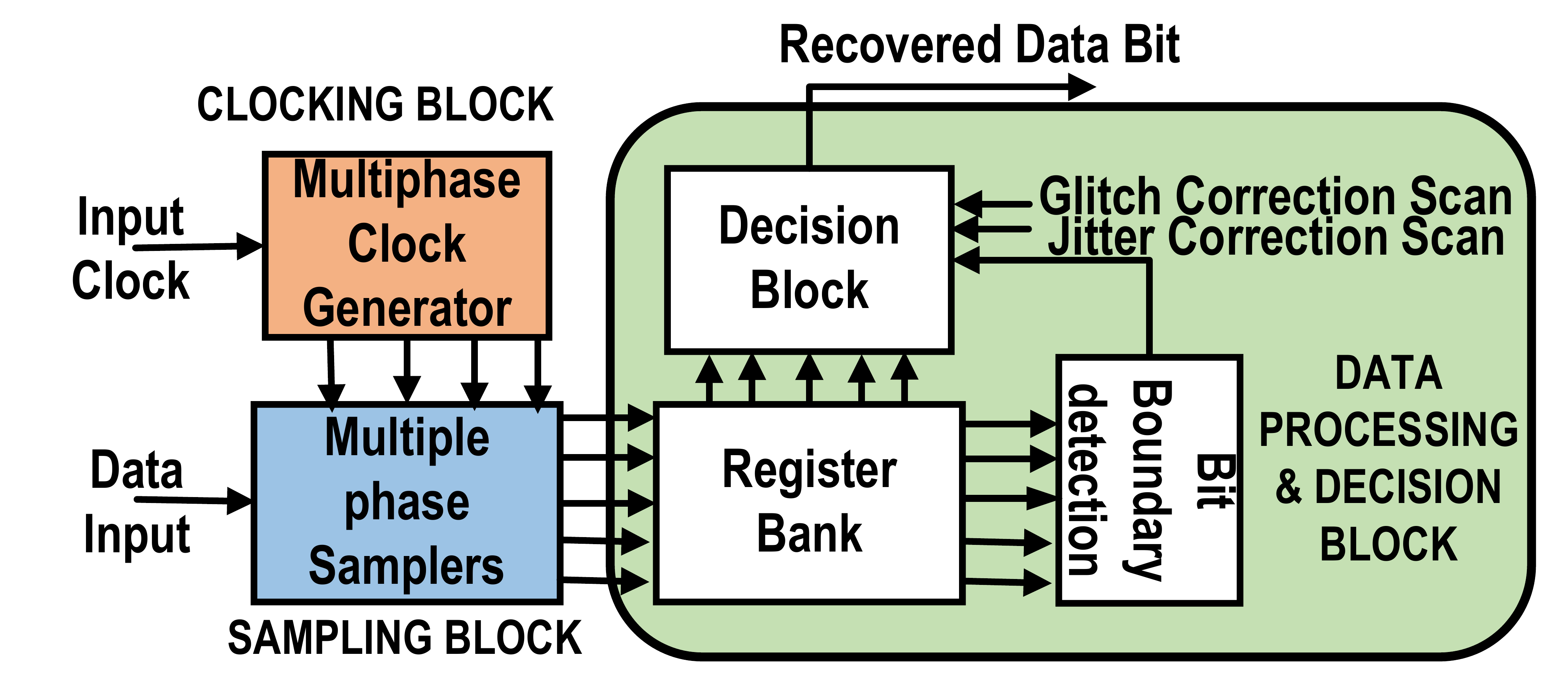}
    \caption{Architecture of implemented Oversampling CDR}
    \label{fig:oversampling}
    \vspace*{-0.2in}
\end{figure}

A fully digital oversampling CDR\cite{OverSamp_Overview} \cite{OverSamp_CDR_Link} is implemented to recover the clocking information. It samples the received data at multiple points, stores them in FIFO registers, before determining the optimal sampling point, which is used to correctly determine the transmitted data. The CDR is also equipped with tunable glitch and jitter correction logic using external scan bits. An external clock is fed to the phase generator block to produce multiple clock phases needed for sampling the input at different points. The sampled data is sent for processing and decision-making, using the predefined logic of the decision block and external glitch and jitter corrections inputs.   Figure \ref{fig:oversampling} shows the block-level architecture of implemented Oversampling CDR. 

\section{Results}
\begin{figure}[!t]
   \centering
    \includegraphics[width=0.48\textwidth]{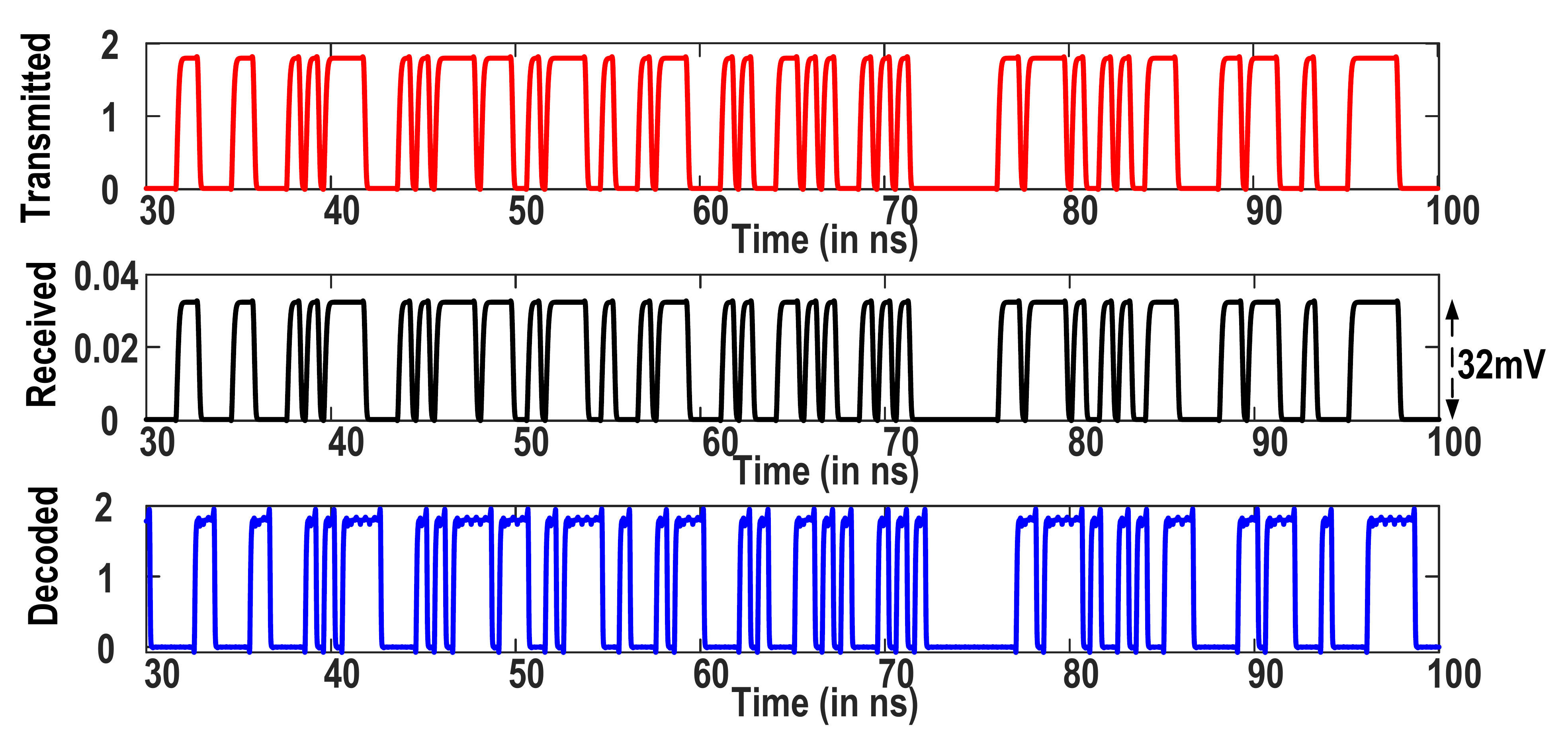}
    \caption{Input/Output Wave-forms at 2Gbps for the SerDes link.}
    \label{fig:SerDes_Timing}
\end{figure}

\begin{figure}[htp]
   \centering
    \includegraphics[width=0.48\textwidth]{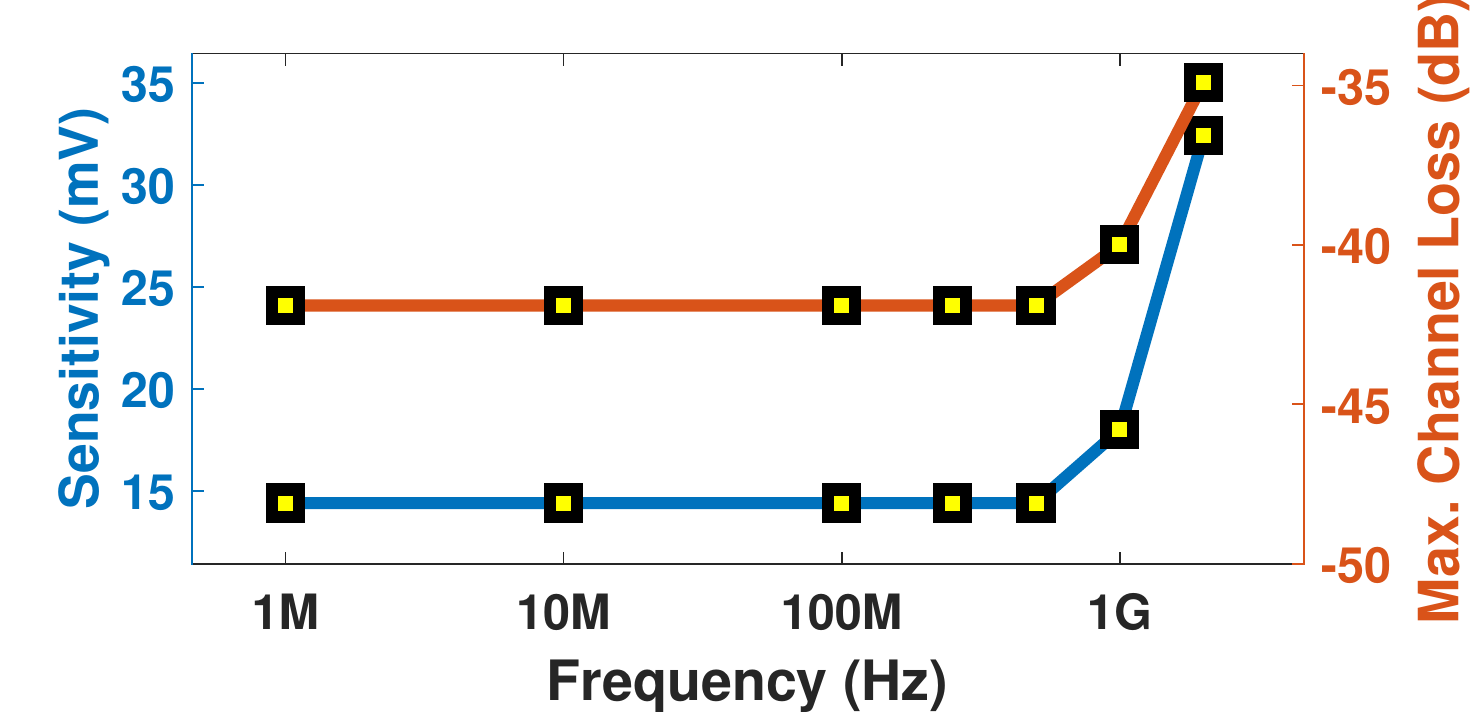}
    \caption{Variation of Sensitivity and maximum operating channel loss of SerDes Receiver with frequency of operation}
    \label{fig:Sensitivity_plot}
\end{figure}

The SerDes design (in GDSII format) is imported in cadence virtuoso to extract parasitics and perform post-layout simulations. Figure \ref{fig:SerDes_Timing} shows the input/output waveforms of SerDes link at 2 GHz with PRBS-31 input. The received signal represents the signal reaching the RX front end, after channel attenuation of 34 dB. 
In contrast with Strong-Arm Latch \cite{SA_razavi} based receiver designs, which are sensitive to even 10 mV signals at 2 GHz, the presented all-digital design achieves $\approx$ 32 mV sensitivity, which is promising in several applications like PCIe interfaces, short-range interconnects, where the required bandwidths are less and channel losses are not significant.
The variation of sensitivity of SerDes receiver with different operating frequencies is depicted in Figure \ref{fig:Sensitivity_plot}. Note that the maximum channel loss corresponding to zero BER transmission is also shown.

\begin{figure}[htp]
   \centering
    \includegraphics[width=0.48\textwidth]{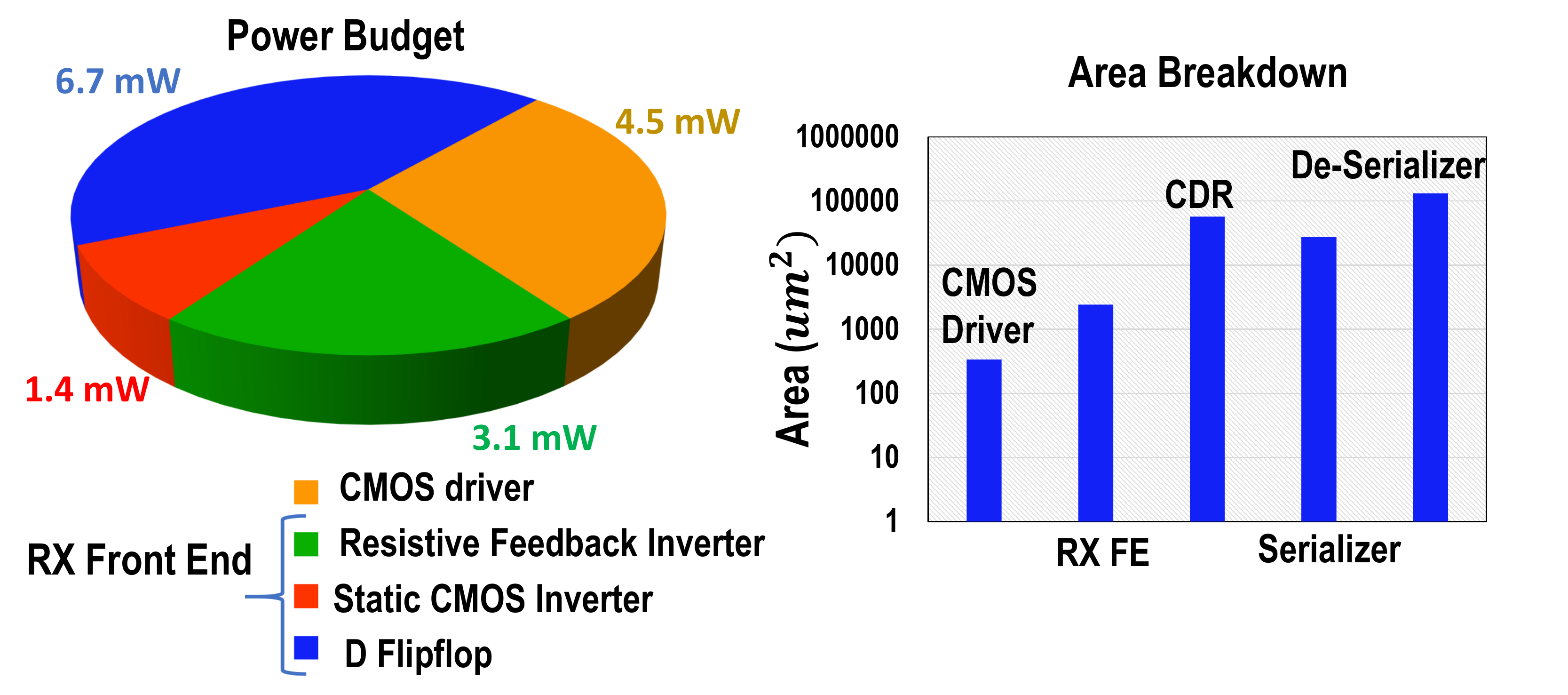}
    \caption{Power Budget and Area Breakdown of the designed Serial link.}
    \label{fig:SerDes_Power}
\end{figure}
\begin{figure}[htp]
   \centering
    \includegraphics[width=0.45\textwidth]{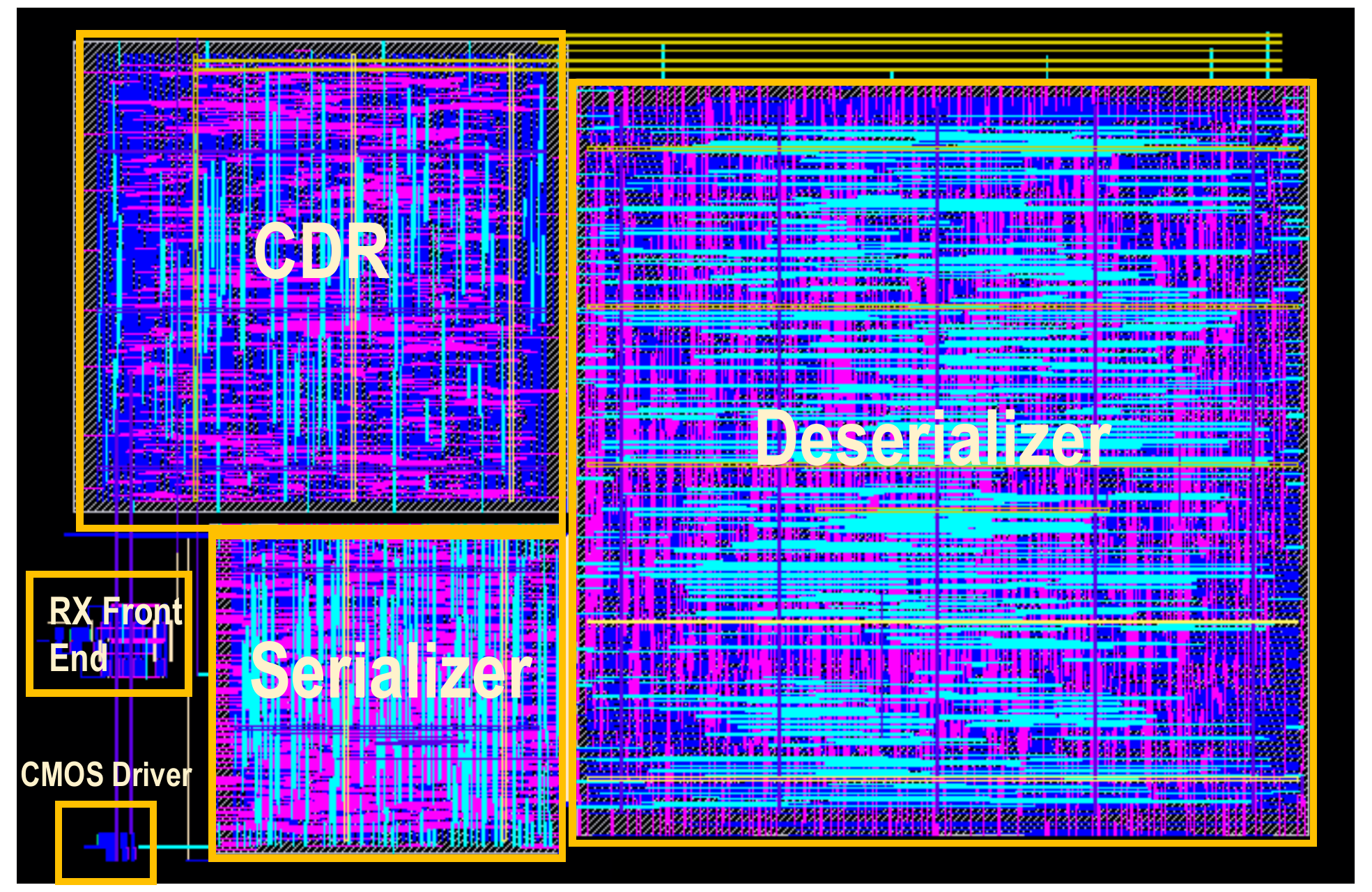}
    \caption{Layout of the generated SerDes. The complete design is available at: \url{https://github.com/SparcLab/OpenSERDES}}
    \label{fig:GDS}
\end{figure}

Figure \ref{fig:SerDes_Power} provides the power consumption  (excluding serializer, deserializer and CDR blocks) and area of the all-digital serial link. The link needs total power of 15.7 mW operating at 2 GHz for the supply of 1.8V, with the receiver and transmitter consuming 11.2 mW, and 4.5 mW respectively. Serializer, Deserializer and CDR circuits require additional 235mW, 128mW and 59mW respectively. The high power numbers of serializer and deserializer blocks are the result of intended design choices to support large IO streams (eight parallel links with 32-bit input) for communication. Thus, the implemented all-digital SerDes operates at 2 GHz for 34 dB channel loss, while consuming power of 437.7 mW, therby exhibiting energy efficiency of 219 pJ/bit.
Figure \ref{fig:GDS} presents the layout of the design. The design layout extends over an area of 0.24$mm^{2}$, with deserializer occupying 60\% of the area, while the CMOS driver and RX Front end cover 0.2\% and 1.1\% respectively.

\section{Discussions}

\paragraph{Pros and Cons of the presented SerDes design} 
The SerDes is equipped with the capability to support multi-GHz transmission, in addition to being a fully synthesizable design, ensuring process-portability. Except for serializer, deserializer and CDR, the entire serial link is implemented using a few CMOS inverters and MOS devices, thereby requiring less chip area compared to mixed signal circuit design, which require additional complex circuitry and I/O pins to provide appropriate biasing for functioning.

However, compared to conventional mixed signal circuit approach, the presented SerDes exhibits slightly lesser sensitivity, and operates at lower speeds, making it appropriate for applications, where the required operating speeds are smaller (1-2 GHz), channel losses are moderate (34dB - 40dB).

\paragraph{Applications} 
The generated SerDes can be used to interface links where the bandwidth requirements is $<$ 2 Gbps, like the PCIe 1.x, PCIe 2.x, PCIe 3.x, PCIe 4.0 bus, where the required Bandwidth/lane ranges from 250 Mbps to 2 Gbps. The SerDes can also find applications in high bandwidth density, short-range interconnects between chiplets, like Intel’s recent high density, multi-die packaging paradigm, Embedded Multi-die interconnects (EMIB) \cite{EMIB_2016},\color{black} where the primary focus is on the area of the serial link, requiring data speeds in the range of 1-4 GHz in a significantly low channel loss setting of 1-5 dB.

\begin{figure}[!t]
   \centering
    \includegraphics[width=0.48\textwidth]{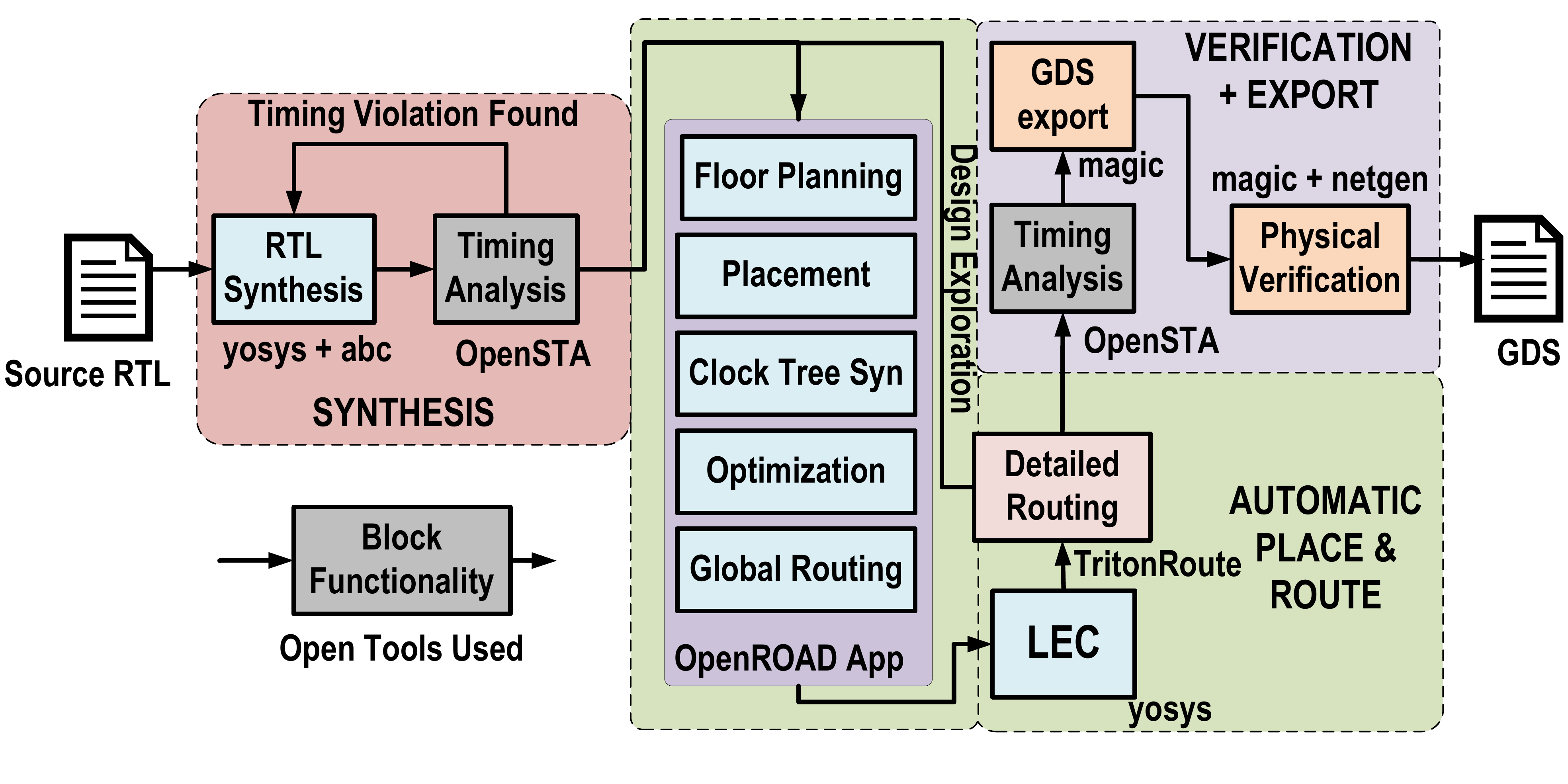}
    \caption{Design Flow of OpenLANE tool}
    \label{fig:OpenLANE_flow}
    \vspace*{-0.2in}
\end{figure}

\paragraph{OpenLANE Tool flow}
The OpenLANE tool\cite{Openlane_repo}, an open source tool with automated RTL to GDS flow, is used to synthesize the RTL and generate GDS for the SerDes design. The OpenLANE tool flow uses custom methodology scripts that perform design exploration and optimization and chose the best implementation, using a number of constituent open source tools, like magic, yosys, OpenSTA, etc., which are invoked at different times in the flow, to execute the Synthesis, APR, Verification and Export steps of the ASIC design flow. Figure \ref{fig:OpenLANE_flow} shows the overview of OpenLANE tool flow, also indicating the open source tools used in each step of the flow.     

\section{Conclusion}
The common roadblocks to hardware/ASIC designs, i.e., the high commitment costs, the much-needed expertise of the designer and the uncertainty (risk) of design functionality, prevent the designers from the required access to advanced technologies. With main objective to democratize the hardware design, the open source hardware revolution is gaining interest, with several open source EDA tools and designs are beginning to surface for public use. These designs and tools can be sourced, modified, or enhanced by the designers, thereby eliminating the involved design cost and the risk factors. The tools are incorporated with machine learning and optimization capabilities, with automated flow, that generate the fully optimized implementation of complex designs. This significantly brings down the applied human effort, and removes the level of needed expertise through automated open source popular EDA tools such as magic, ngspice, OpenRoad, OpenLANE, BAG etc. The advent of open source PDK by Google and Skywater technologies has provided designers with free access to the CMOS 130 nm process, equipping them with standard library needed for generating designs. Thus, Open source hardware significantly reduces the development time and costs for the design.
Technology scaling is also shifting the focus to digital-only designs as analog circuits suffer from short channel effects and degraded performance at low voltage supplies. All-Digital or synthesizable designs also enable faster time-to-market than analog designs.
With the development of APR tools, the all-digital designs can be easily ported to new process with minimal effort. 
Consequently, all-digital designs are being preferred today over analog or mixed signal designs.

The work presents the first open source SerDes using Skywater 130 nm OpenPDK, aiming to build a synthesizable design, ensuring process-portability, without compromising with the performance of the design. Simulation results show that the designed SerDes operates at speed of 2 GHz, has a receiver sensitivity of 32 mV to cope up with the channel losses of 34 dB, while consuming power of 438 mW. The generated SerDes finds applications in interfaces, where the required data rates are not very high, for example in PCIe bus interfaces and in short-reach interconnects between chiplets. The data speed and power is primarily limited by the Skywater 130 nm, the only Open PDK available presently, and will improve significantly as new OpenPDK's are made available.
The presented work signifies the beginning of the open source ASIC designing, especially in traditionally mixed-signal-heavy circuits such as the SerDes, and may serve as a guide to prospective hardware/ASIC designers, illustrating the importance of open source hardware, and motivating them towards digital-friendly designs using open source tools and PDK.

\section*{Acknowledgment}

This work was supported in part by the Semiconductor Research Corporation (SRC) CHIRP Program, under Grant No. 2878.014 and the National Science Foundation (NSF) Career Award under Grant No. ECCS 1944602. The authors would like to thank Nirmoy Modak and Debayan Das, Ph.D. students at Purdue University for their valuable feedback during writing the manuscript.

\bibliographystyle{unsrt}
{
    \bibliography{mybibfile.bib}
}


\vspace{12pt}
\end{document}